\documentclass[8pt,a4paper,twocolumn]{article}
\usepackage[utf8]{inputenc}
\usepackage[T1]{fontenc}
\usepackage[french, english]{babel}

\usepackage{graphicx}
\usepackage{caption}
\usepackage{amsmath,amsfonts,amssymb}
\usepackage{abstract}
\usepackage{subcaption}

\title{Modeling and optimizing a distributed power network : A complex system approach of the prosumer management in the smart grid. \\ Survey}
\author{Nicolas Gensollen, Vincent Gauthier, Michel Marot, Monique Becker}

\begin{document}
\twocolumn[
\begin{@twocolumnfalse}
\maketitle
\begin{abstract}
\begin{small}
\textbf{One of the most important goals of the 21st century is to change radically the way our society produces and distributes energy. This broad objective embodies in the smart grid’s futuristic vision of a completely decentralized system powered by renewable plants. Imagine indeed such a real time power network in which everyone could be a consumer or a producer. Based on a coupled information system, each user would be able to buy/sell energy at a time depending price that would allow a homogenization of the consumption, eradicating the well-known morning/evening peak. This attractive idea is currently booming in the scientific community as it generates intellectual challenges in various domains.}

\textbf{Nevertheless, lots of unanswered questions remain. The first steps are currently accomplished with the appearance of smart meters or the development of more efficient energy storage devices. However, the design of the decentralized information system of the smart grid, which will have to deal with huge amounts of sensor’s data in order to control the system within its stability region, seems to be still in search.}

\textbf{In the following survey, we concentrate on the telecommunication part of the smart grid system. We begin by identifying different control level in the system, and we focus on high control levels, which are commonly attributed to the information system. We then define a few concepts of the smart grid and present some interesting approaches using models from the complex system theory. In the last part, we review ongoing works aiming at establishing telecommunication requirements for smart grid applications, and underline the necessity of building accountable models for testing these values.} 

\end{small}

\end{abstract}
 \end{@twocolumnfalse}
]

\tableofcontents 

\section{Control of Power grids}
Having a reliable access to electricity is nowadays taken for granted and consumers expect high quality power at any time and at any desired amount. However, supplying power to millions of customers (EDF has around 27.7 million customers in France) turns out to be a delicate know-how. Actually, it is known that electricity is a hardly manageable quantity. Its delicate storage, its degradation during transport, as well as its almost non routability are only a few examples of the difficulties of handling power compared to data for instance.

Nevertheless, catastrophic events like blackouts or outages are relatively rare and localized, meaning that, despite the underlying difficulties, power operators found solutions to make power networks reliable and more or less efficient. Countries have built over the years their own power grids with their own actors and processus. Thus, our interest in this section does not lie in studying each of them in details, but rather in understanding how current power networks globally function, how resilient they are, or which typical weaknesses they reveal.

The geographical position of France, at the crossroads of western Europe, as well as the monopolistic past of its operator (EDF : \textit{Electricity of France}), make the french grid example really interresting. Since 2000, RTE, a subsidiary of EDF is in charge of the electricity transport and acts toward a safer and more stable grid. Its website ($ http://clients.rte-france.com $) provide a large amount of explanations and data (consumption predictions, losses, margins...).

\begin{figure}[h!]
\centering
\includegraphics[scale=0.4]{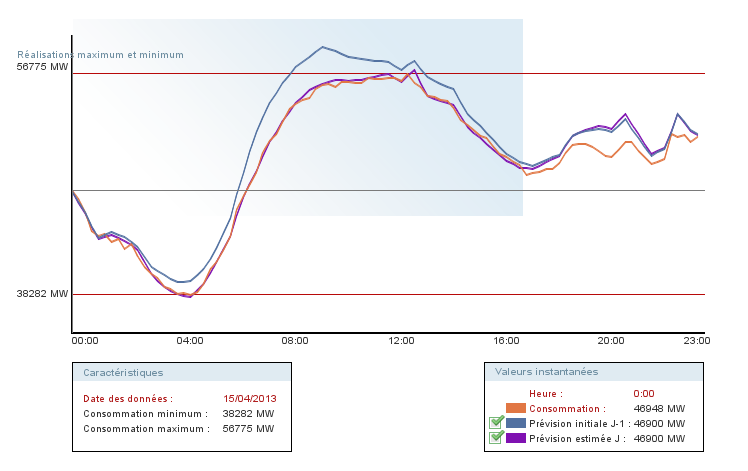}
\footnotesize{ \caption{French consumption (real (red curve) and forecasted a day before (purple curve) and the given day (blue curve)) the 15/04/2013 (source : RTE)\label{fig:conso}}}
\end{figure}

\begin{figure}[h!]
\centering
\includegraphics[scale=0.5]{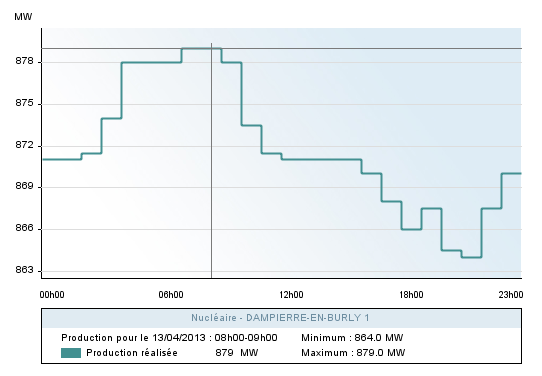}
\footnotesize{ \caption{Hourly production of the reactor 1 of Dampierre-en-Burly nuclear power plant, the 13/04/2013 (source : RTE)\label{fig:prod}}}
\end{figure}

Energy production is nowadays very centralized and mainly achieved through 58 reactors spread accross 19 nuclear power plants in the case of France. Other production plants such as hydraulic or gaz are also used to a lesser extent. A daily production schedule for each reactor (cf. figure \ref{fig:prod}) is computed based on a combination of data studies (such as consumption forecasts (see figure \ref{fig:conso})) and power flows algorithms, and is adjusted regularly.

\begin{figure}[h!]
\centering
\includegraphics[scale=0.4]{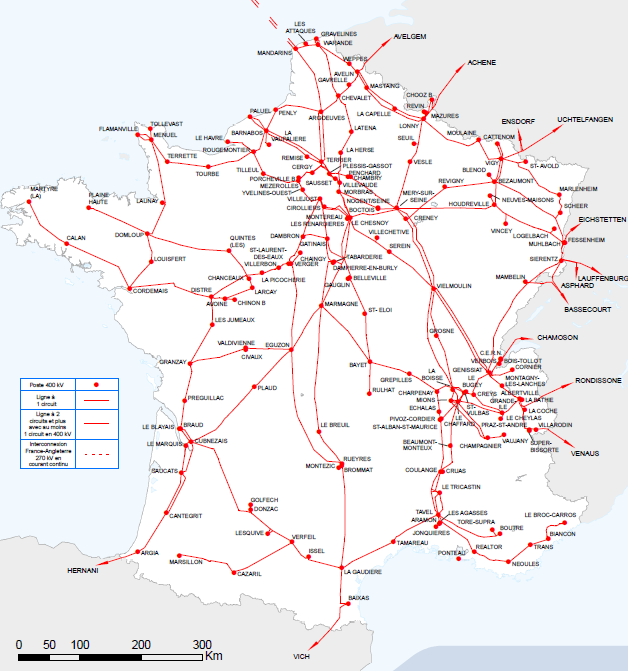}
\footnotesize{ \caption{French High voltage transport network (source : RTE) \label{fig:HT}}}
\end{figure}

\begin{figure}[h!]
\centering
\includegraphics[scale=0.7]{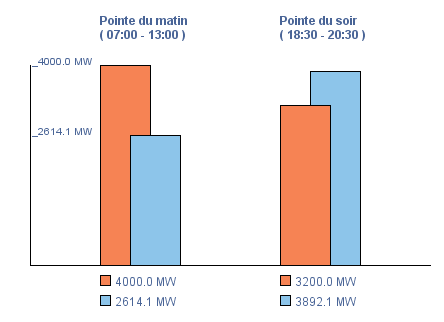}
\footnotesize{ \caption{Required Margin for stability (orange) and available margin after allocation (blue) for morning and evening peacks the 09/02/2012 during an extremely cold week which induced the largest energy consumption ever recorded in France 4 days in a row. This example is extremely rare as the required margin for stability was not met in the morning, meaning that the grid was operating on the edge (source : RTE) \label{fig:margin}}}
\end{figure}

Once producted, electricity is converted by transformers to very high voltage (400 kV) in order to transport it on long distances (cf figure \ref{fig:HT}). This power grid, called (very) high voltage grid, is, in France, managed by RTE. It is important to note that this network is interconnected to neighboring countries in order to exchange energy in real time (see $ http://www.epexspot.com/en/ $ for instance). The stability of this network is absolutely paramount, and is maintained by establishing production margins (cf. figure \ref{fig:margin}), meaning that the production is scheduled volontarely higher than forecasted demand in order to cope with contingencies. Nevertheless, interconnections with other countries as well as partnerships with secondary actors (called \textit{responsables d'équilibre}, which means "\textit{stability officials}") are common ways to maintain stability as energy can be sold on international markets, or injected/withdrawn by these stability actors.

Transformers (around 2350 accross the french high voltage network (100 000 km of lines) for instance) are responsible for adjusting the power voltage between networks :
\begin{itemize}
\item Very High voltage (400 000 V or 225 000 V)
\item High Voltage (90 000 V or 63 000 V)
\item Medium voltage (20 000 V)
\item Low Voltage (400 V or 230 V)
\end{itemize}
700 000 transformers interconnect the 586 000 km of medium voltage lines to the 654 000 km of low voltage lines. A very large majority of customers are located in low voltage grids and do not participate in the maintenance of the grid. Customers of a same region (downstream of a transformer) are typically considered as an aggregated load in simulations. Some industrial customers (transportation companies (RATP, SNCF) for instance) are however directly connected to high/medium voltage grids as their needs greatly exceed regular users'.

Stability actors like RTE need communication means between generating plants, control sites, or other stability operators... A classical solution used by RTE consists in deploying optical fiber along electrical lines through various techniques. Some of them are actually really effective as portions of 10 km could be installed within a single day. This has a few consequences such as a same topology for high voltage grid and data network, meaning also that the fall of a line causes the fall of a data link. Actually, one of the questions in smart grid revolves around determining robust topologies for electrical grids and their coupled communication networks.  

In theory, in high and very high voltage grids, redundancy and structural robustness (mesh) are introduced while most of low voltage grids have kind of a tree/star structure. This is easily understandable and common to data networks for instance. Later in this survey, topological properties of power grids will be discussed in more details. Nevertheless, the optimal smart grid topology in terms of robustness, economical cost, and feasability remains still unclear.

This section provided only a little overview of the global structures and control means of the current power grids. The ways operators such as RTE daily deal with stability problems and exchange energy on international markets are a wide scientific field far beyond the scope of this survey, and more information can be found in litterature \cite{Foundation1996} \cite{RTE}. 

\section{Toward a new generation power grid}
Despite functioning, our current way of powering the grid based on fossil energies is becoming delicate as these are becoming scarcer and scarcer while the demand keeps progressing over the years \cite{Ajmone-Marsan2012}. A very nice way of resolving this problem as well as providing a cleaner power system is often assumed to result from the integration of numerous renewable plants inside the grid. These “multi-behavioral micro-generators”, whose purpose is to be as close as possible to the end users, will populate the low voltage grid and deliver energy depending on external parameters such as meteorological conditions for instance. This is a very delicate point as these generators, in addition of being numerous, exhibit also uncertain generation profiles and can hardly be scheduled over time. This means that current power engineering philosophy of estimating the demand and scheduling the production in accordance (cf. previous section) may not provide a proper control for such a dynamical and decentralized system. Furthermore, assuming also the development of electrical cars and storage devices, the structure of this futuristic system seems to be really different from our current hierarchical and centralized power network. How can we then determine the system’s characteristics (topology, control…) that will provide the necessary resilience for the smart grid to become reasonably responsible for our energy alimentation ? 

The recent catastrophic blackout events (Italy 2003, North America 2003, etc…) \cite{Rosas-Casals2007} \cite{Schneider} remind us indeed that, no matter how robust the power grid seems to be, small events, like failures of components, can potentially trigger large wide-spread cascades \cite{Sachtjen2000}. Failure events are hardly predictable as they may result from extremely various events (storm, heat, dysfunctions…), and the classical way of dealing with them is to introduce some redundancy in the system so that failure of a component does not trigger cascades.

Nevertheless, even without failure, the power grid remains a dynamical and non linear system, which means that small perturbations like local loss of synchrony can be amplified and cause the desynchronization of the whole system. In other words, the power grid has to be driven continually within its stability basin (in terms of respect to the grid’s main frequency for example). This task requires some synchronization among the grid’s devices as well as the possibility to drive them without deviating too much from this synchronous state \cite{Dorfler2013}. Given the demand forecasts, one of the operator’s main concerns lies then on the power production allocation among the generators, in order to meet the demand while staying as close as possible to the synchronous stable state \cite{Foundation1996}. 

These centralized control operations are well established for the current power networks (cf. previous section). However, this approach has to be reconsidered in a scenario with decentralized power production system. Especially with a high penetration of renewable sources that generate power according to meteorological conditions that can only be vaguely estimated \cite{Budischak2012} \cite{Gast2012}, the stability questions as well as the insurance to meet demand become critical.
 
 \section{Control layers approach}
Studying the control of a complex and heterogeneous system such as the smart grid is not a trivial task. One of the reasons of assigning a communication system to the power grid is indeed to guarantee a better and finer control of the system. Does it mean that the communication system will be responsible for all the control operations within the system ? If so, would a best effort type of communication network be able to deal with power stability questions with milliseconds timescales ? 

Actually, it is common to visualize the control operations in power networks on distinct layers based on their timescales \cite{Bolognani2011}. Basically, low layers handle security and quality of power while high level layers deal with market or large scale prediction operations. Scale and time constraints are thus really different from one layer to another, which suggest that the control system should also be designed with a layer approach.

Primary control algorithms are executed locally, and deal with local power stability. The use of a communication network seems thus unnecessary and stability is ensured by electrical processes such as droop voltage control on inverters for instance \cite{DeBrabandere2004} \cite{Engler2005}. In other words, this level of control remains really close to electricity and need thus to be designed while taking into account complex phenomenons due to non-linearities.

Higher control algorithms receive a certain number of constraints from lower levels in order to guarantee "lower stabilities". The second level receives for instance contraints from the first level, and should take care of optimizing the network operations within these contraints \cite{DeBrabandere2007}. In other words, high level algorithms do not worry about power quality as it is supposed to be taken care of by lower level, they are rather concerned about sending appropriate commands to the right generating plants. The necessity of a communication network is thus clearly apparent and cause a number of issues that will be tackled along the present survey.

However, as briefly explained in the introduction, the idea behind the smart grid concept is not only a better controlled power grid, but a dynamic, decentralized, resilient system, whose adaptability is incommensurate with the current power grid's. In this vision, the system should allow a vaste number of units behaviors and deal with uncertainties while being stable and resilient.

The following section introduces new concepts that are necessary in order to properly understand smart grid's remaining challenges.

\section{High level concepts and challenges}
\subsection{Demand-Side Management}
Current philosophy is that power generation should adapt to power consumption. That is, consumers don't take into account anything else than their own energy needs. One of the results is the well-known and very expensive energy peaks, which induce the use of more polluting resources in order to meet the demand.
The will of eradicating, or at least reducing, these peaks is already quite old, and a few principles have been already established. The demand side management (DSM) consists in providing the grid with technologies that make consumers more responsible toward their consumptions \cite{Ramchurn}.

A first step toward demand side management was the introduction of time depending pricing, which divides days in time slots with different prices. The main reason behind this being that consumers are supposed to adapt their consumptions to reduce the bills.

Real-time pricing (RTP) is in the continuity with a finer slots decomposition, and a price that is adaptable in accordance with production and transport conditions \cite{Ramchurn2011}. Nevertheless, as consumers are not always looking and reacting at energy prices to save money, the idea of using RTP for the smart grid is a little bit different and lies on the developpement of advanced metering infrastructures (AMI).

Advanced metering infrastructures (AMI), such as smart meters for instance, are indeed responsible for collecting information from the grid (such as real-time prices) and acting accordingly. AMI could for instance decide to delay some appliances if a sudden incident on the grid causes a price explosion. Obviously, AMI should make the difference between delayable appliances (air conditionner, batteries charging...) and non delayable appliances (fridge, lights...), and allow consumers to indicate their preferences.

With the development of advanced metering infrastructures, real-time price stands for a control parameter of the consumption, giving to the end users a condensed information of the network state and production conditions. A real continuous time depending price is addressed in a few publications, but dealing with such an information is rather complex and relatively unrealistic for the moment. Actually, continuous price causes difficulties on several layers. Imagine indeed the quantity of information that has to be exchanged for every appliances to know at any time an information that has to be computed over a large number of measures. Similarly, imagine receiving at the end of the month a very expensive electricity bill because your air conditioner operated in periods where prices where very unstable with a succession of peaks, or the same air conditioner continuously turning on and off because the electricity price is unstable. 

Thus, most of publications consider a discrete pricing model where a price is computed for the next period (hour(s), day(s)) and communicated to end users. With such an information, smart appliances can decide weather to operate or not, that is, weather the utility of operating is higher than the resulting price of operation (cf. next subsection). Based on machine learning algorithms (price, demand patterns, user habits) and optimization, smart meters can then organize appliances operations in the most efficient ways. 

However, some publications addressed a few negative consequences of such a system \cite{Boudec2012} \cite{Ramchurn2011}. An easily understandable one is the appearance of secondary consumption peaks at unusual periods (ends and beginings of low price periods, or very low price periods for instance). Smart appliances waiting until the end of expensive time periods for operating, or users doing a maximum operations just before the end of low price periods result in secondary consumption peaks that can be also very expensive for the network operators.

Another subtile consequence is a negative evaporation of the frustrated demand \cite{Boudec2012}. That is, the demand that cannot be met at a time t due to various production or transport reasons is amplified at time $ t+ \Delta t $. Consider a given house heaters for example, that decide to stop operating for a few hours due to higher electricity prices during this period. The amount of energy needed after a few hours to warm the house at the desired temperature might be higher than the amount of energy that would have been needed to maintain a constant temperature the whole time, meaning that not satisfying the demand at a time t can results in energy peaks at futur moments. It can be shown that this kind of system is unstable \cite{Boudec2012} and thus, doest not provide a proper demand-side management mechanism.

The domain of appliances optimization under price constraints exhibits similarities with domotic or internet of things for instance and generates a lot of questions and challenges way beyond the scope of the present survey. Nevertheless, as will be addressed in the next subsections, the smart grid complexity does not stop downstream of the smart meters.

\subsection{Prosumers and Virtual Power Plants (VPP)}
Actually, the smart grid's golden idea revolves around the fact that everyone could behave as a consumer or as a producer if he has the capacity of doing so. This new agent model, often called \textit{prosumer}, aims at defining a more realistic behavioral model and allowing more complex actions that the blindless consumption of the needed energy \cite{Rathnayaka2012}.

Prosumers are actually economically motivated in the sense that they try to find the optimal compromise between outlays and consumption, or try to maximize their profits when they are selling energy \cite{Samadi2012} \cite{Rathnayaka2012} \cite{Grijalva2012}. Here is indeed the complicated part because allowing almost any agent to inject energy in the system results in a huge amount of questions and problems (technical feasability, stability, power quality injections...). The ways and consequences of allowing such behaviors are far beyond the scope of this paper, nevertheless, it is obvious that a single prosumer alone will probably never reach the grid's requirements (in terms of power quality or stability for instance), and will thus never be allowed to inject energy inside the grid. 

Partly due to this observation, the virtual power plant (VPP) concept has been imagined and can be thought of as an aggregation of distributed generators and loads dispersed among the network, but controllable as a whole generating system \cite{Braun} \cite{Member2011}. This interesting idea dematerializes the energy production and envision a giant market cloud where distributed generators can form virtual economical clusters in order to gain visibility and maximize their profits. These virtual aggregations based on economical reasons are often termed to as Commercial Virtual Power Plants (CVPP).
A little less abstract are the Technical Virtual Power Plants (TVPP) concept, which is also an aggregation of distributed units, but within a same geographic location. Basically, CVPP aggregates distributed generators based on profits maximizations and provide then information to the TVPP, which are active on the distribution level. The virtual power plants system architecture is a quite new field of research, and more information could be found in \cite{Braun} \cite{Member2011} \cite{Technique2003}.

The prosumer model is thus relatively complex since these agents could behave as typical consumers with different profiles and more or less developped ecological sensibilities, as well as coalitional producers that look for maximisation of profits by forming alliances. As explained previously, it is not in the present paper's scope to deal with all questions that such a model raises. However, such complex and dynamical coalitions need obviously communication means in order to function properly. In other words, the feasability of these concepts have to be tested on multiple models with regards towards the different part of the smart grid infrastructure. A question of interest is then to understand what are the telecommunication requirements for implementing these concepts and confronting them to the reality of data networks. 

A common way of modeling these behaviors comes from the micro-economic utility functions and the game theory. In this direction, \cite{Samadi2012} proposes for instance a Vickrey-Clarke-Groves (VCG) mechanism aiming at maximizing the so-called social welfare (basically, the aggregate utility functions of all users minus the total cost of energy). Typical utility functions for aggregate load are quadratic functions corresponding to linearly decreasing marginal benefit of the form ($ \alpha $ being a predetermined parameter) :

\[ U(x,\omega) = \left\{ \begin{array}{ll}
                                        \omega x - \dfrac{\alpha}{2}x^{2},\ if\ 0\leqslant x \leqslant \dfrac{\omega}{\alpha} \\
                                        \dfrac{\omega^{2}}{2 \alpha},\ \ \ \ \ \ \ \ \ \ if\ x\geqslant \dfrac{\omega}{\alpha} 
                                        \end{array}
                                         \right.
                                         \]
Under a few reasonable assumption (cf. \cite{Samadi2012}), quadratic functions are also usefull to account for the cost $ C_{k} $ of providing $ L_{k} $ units of energy at time slot k : 

\[C_{k}(L_{k}) = a_{k}L_{k}^{2}+b_{k}L_{k}+c_{k} \]

$ a_{k},\ b_{k},\ and\ c_{k} $ are fixed positive parameters that depend on the time slot considered.

\begin{figure}[ht]
\centering
\includegraphics[scale=0.5]{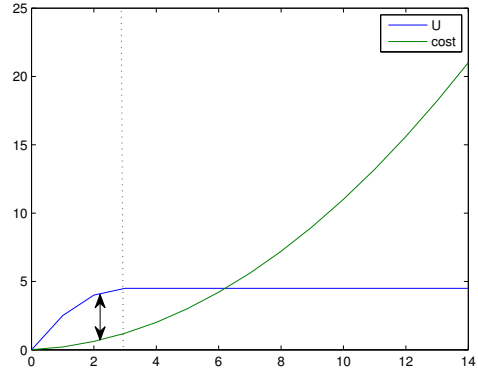}
\footnotesize{ \caption{Example of a prosumer(s) behavioral model : $ \omega = 3 $, $ \alpha=1 $, $ c_{k}=0.01 $, $ b_{k}=a_{k}=0.1 $. The objectiv of the agent(s) is the maximization of his utility (or the sum of their utilities) minus the resulting energy cost. Thus, the region to the right of the dashed line is non interesting for the consumer(s) as his utility is already maximum. \label{fig:cost}}}
\end{figure}

An efficient allocation consists then in maximizing the sum of the utility functions of all users while the cost for the energy provider is minimized (cf. figure \ref{fig:cost}).

The formalism of utility/cost functions as well as game theory (competitive equilibrium, Nash equilibrium...) can be deepened in scientific litterature \cite{neumann44a} \cite{Samadi2012}. We believe that it could provide interesting tools (see \cite{Kallitsis2010} for instance) in order to model single unit's behaviors as well as clusters of units such as VPP for example.

\subsection{Microgrids and Islanding}
Another principle of the smart grid is that generation and consumption are geographically closer than in the current power grids, meaning that transport losses could be diminished. Considering a high level of distributed generation penetration, the end part of the grid will be populated by small generators, storage devices, and variable loads. A microgrid can thus be seen as a localized group of generation, storage, and load units that operates with more or less support from the main grid. Actually microgrids' inner units can during certain periods of time meet their own needs without any support from the main grid, making the microgrid self-sufficient \cite{Erol-kantarci2011}.

The islanding concept results logically from the previous observation. Islanding basically means disconnecting microgrids from the main grid, which can be the resulting procedure when a problem occurs in the main grid (stability issues, line outages...). In other words, the islanding concept enables the disconnection of a given number of self-sufficient microgrids in order to relieve the main grid, and can thus be seen as a way of containing cascades of failures. This can also be seen the other way around as self-sufficient microgrids can island themselves in order to avoid being hit by failures spreading from the main grid.

Furthermore, if a microgrid is not only self-sufficient but also produces more energy that it actually consumes, it has a few options such as forming or joining a VPP (cf. previous section) in order to sell the excess, or it can fill storage devices for instance. However, in a delicate situation where the main grid enconters unstabilities and need to be relieved, overproducting microgrids could associate with underproducting ones in order to form globally self-sufficient clusters of microgrids. This yet very theoritical concept, often called super-islanding, is adressed in a few publications dealing mainly with cascades phenomenons in smart grid systems \cite{Pahwa} \cite{Negeri2012}. Super-islands are, even so, different from VPP as the purpose revolves around avoiding failures cascades, and not around forming economically profitable virtual coalitions.

Actually, in a line failure spreading context, power grids tend to loose unintentionally connectivity, resulting in clusters disconnected from the main grid, meaning clusters of users without energy \cite{Rosas-Casals2007}. Sometimes, only very small failures result in this kind of connectivity loss, which in a centralized power grid can easily result in localized blackouts. A theoritical strength of distributed system is indeed to maintain operation even in a disconnection scenario. We can thus expect smart grid systems, which enable islanding processes, to exhibit better resilience than their centralized "dumb" counterparts.

The spreading of cascades in networks and in interdependant networks is indeed a very popular field of research in the complex system theory \cite{Watts2002} \cite{Gao2011} \cite{Leicht2009} \cite{Shao2010} \cite{Buldyrev2009} \cite{Brummitt2012}. Numerous publications study how power grids across the world react when a certain fraction of devices fail, and try to find links between topology and resilience \cite{Rosas-Casals2007} \cite{Nakarado} \cite{Wang2009}. The widely-used percolation approach for studying the evolution of the connected component's size enables a visualization of the effects of a cascade \cite{Leicht2009}. Based on these tools, it has been shown that islanding could theorically reduce drastically the size of cascades.

Islanding and microgrids are important concepts in the smart grid system, providing it more resilience and adaptability against failures and unstabilities. Nevertheless, if these concepts exhibit interresting and encouraging results in mathematical models, their feasabilty regarding real networks' operating conditions as well as their requirements appear to need deepened researches on more realistic models. 
\begin{quote}{ \textbf{D’Souza} : }
\textit{"We have to start from the data-driven, engineered world and come up with the mathematical models that capture the real systems instead of using models because they are pretty and analytically tractable."}
\end{quote}

\subsection{Challenges}
The previous subsections presented some of the main conceptual ideas of the smart grid's overall system. Theoretically interresting, they also provide challenging research orientations in order to adapt them to reality constraints. Demand side management, for instance, starts with the proper design of advanced metering infrastructures (AMI) such as sensors or smart meters as these are the basic components on which the global architecture will rely on. Regarding the design of these AMI, engineering questions come from various fields (electronic, telecommunications, design...) and are more or less interdependant. In other words, building such an architecture revolves around compromises between seemingly incompatible things, meaning that one should be aware of limiting values beyond which operation becomes impossible. A sensor for instance could be designed to consume as little energy as possible while needing a large emission rate which in turns requires energy. Knowing limiting operating values for this sensor's energy consumption and emission rate appears thus indispensable. However this has to be in accordance with the rest of the system (the specifications of the upstream smart meter which deals with these sensors information for instance), which suggest that these values of key parameters should be tested on simulations based on realistic models. One of the key challenges for the smart grid when it comes to telecomunications (but not only) revolves then around developing simulation tools. This approach is already undertook by several serious institutions such as the National Institute of Standards and Technonoly (NIST, USA) and networks' requirements per smart grid applications are studied and questionned. 

Overall, challenges in smart grid are wide and more or less precise. In \cite{Ramchurn}, the author decorticates the smart grid global problem and extricates crucial challenges for the information system in key components of the project (Demand-Side management, Electric vehicles, Virtual power plants, Prosumers, and Self-healing networks). This includes complex search and optimization algorithms as well as distributed autonomous agents that can take decisions and act without a centralized control unit.

\section{Complex systems and Consensus}
\subsection{Using complex system theory for smart grid modeling}

Studying large systems of interracting units by modeling and simulations is classic in the complex systems' field \cite{Newmanc} \cite{Boccaletti2006} \cite{Cohen2007}. Actually this discipline sees currently a lot of interest from researchers from various domains, as its tools enable studying statistically several conceptual properties (robustness, dynamics...) of large systems \cite{Gai2010} \cite{Buhl2004} \cite{Strogatz2005}. As a matter of fact, power engineering does not make an exception, and power networks have been studied extensively with this approach for a few years. In fact, it appears that large blackout events have the consequence of stimulating publications in this area as links between resilience and other properties (topology, synchronization propensity...) are carefully sought \cite{Rosas-Casals2007} \cite{Nakarado} \cite{Wang2009} \cite{Chassin2005}.

Basically, pure topological studies generally consider the power grid as a simple undirected graph G defined as a pair (V,E), V being the set of vertices (generators, loads, transformers...) and E the set of edges materializing the electrical lines between the vertices. Assuming that the set of vertices (also called nodes) is organized, it is possible to define the so-called adjacency matrix in the following manner : \[ A_{ij} = \left\{ \begin{array}{lll} 
                         1\ if\ there\ is\ an\ edge\ between\ i\ and\ j \\ 
                         0\ otherwise 
                      \end{array} 
              \right. \] 
This matrix is really important in graph theory since it translates the graph topology in a single mathematical entity. A whole part of the complex systems theory is indeed based on the spectral analysis of matrices inspired of A, the most famous one probably being the Laplacian matrix : $\Lambda = D-A$ with $ D_{ii}=\sum_{j}a_{ij}=k_{i} $ (see \cite{Merris1994} for a complete survey on Laplacian properties). 

As the number of vertices grows, other convenient ways of describing graphs are adopted. The degree distribution P(k) for instance, that gives the probability that a given node has a degree k is indeed commonly used and its form is even a traditional way to classify graphs. A scale-free network for example posesses a distribution of vertex degrees heavily right skewed (power law $ P(k)\sim k^{-\gamma} $) with $ \gamma $ generally between 2 and 3 in real networks. In other words, this kind of networks has the particularity of possessing a few very well connected hubs as well as a large majority of poor connected nodes. This is for example the case of Internet or the World Wide Web (Google or Yahoo are in this case good examples of hubs). 

Publications on power grids' topologies often state an exponential degree distribution (sometimes also a degenerated power law), which indicates the virtual absence of high degree nodes (also called hubs) in power grids. Unlike scale-free networks, which are extremely fragile against targeted hub-oriented attacks (imagine the consequences of google falling for instance), this rather suggests that attacking power grids based on node degree will probably not be effective.

Actually, a few publications studying cascades on power networks suggest that the degree measures may not be relevant to describe the dynamic nature of power grids \cite{Hines2010} \cite{Brandes2005}. Continuing in this direction, \cite{Hines2006} shows that, based on betweeness centrality measures, power grids exhibit power laws distributions, meaning that, despite of having "\textit{degree hubs}", power grids could exhibit "\textit{electrical hubs}". Furthermore, this apparent robustness against targetted attacks is thus seriously questionned as attacking electrical hubs provide a relatively better strategy than random or degree-oriented attacks. As a matter of fact, abstracting power networks in a mathematical undirected graph with shortest path metrics only provide, at best, topological studies electrical circuits. Indeed, once injected in the network, electricity splits along branches due to complex impedances relations and only obeys Kirchhoff's laws. In other words, pure topological models do not necessary provide insightful results when applied to power grids and other ways of dealing with their dynamical nature should likely be explored.

\subsection{Synchronization}
Previous topological approaches tried to find statistical properties on large scale simplified models of power grids. Tackling the problem the other way around, a few publications start their reasoning at a single line connecting two nodes \cite{Filatrella2008} \cite{Dorfler2013} \cite{Motter2013}. Considering indeed a generator feeding a rotating machine through an electrical line, it is possible to express the power transmitted from the generator to the load as a function of the voltage phase angle difference between these two entities (cf. \cite{DeBrabandere2004}) :
 \[P_{ij}(t)=P_{ij,MAX}sin(\theta_{i}(t)-\theta_{j}(t))\]
Where $ P_{ij,MAX} $ is the maximum power flow between i and j, and $ \theta_{i}(t) $ and $ \theta_{j}(t) $ are the voltage phase angles of i and j at time t. 

In other words, there is a link between node variables $ \theta_{i}'s $ and power flows. This is very interresting when making an analogy with synchronization studies in networks of interacting units. 

Natural phenomenons like a swarm of fireflies synchronizing their flashing frequencies, the self-synchronization of coupled oscillators \cite{Acebr}, or the oscillations of a bridge due to the synchronization of pedestrians steps \cite{Strogatz2005} are indeed well studied by researchers of complex systems theory. Modeling the behavior of each entity as \textit{dynamic of i} = \textit{self-behavior of i} + \textit{interactions}, a question of interest consists in seeking links between the synchronization and the underlying topology representing the strength of the interractions.

Providing a usefull tool for this purpose, the Kuramoto model \cite{Acebr} \cite{Boccaletti2006} assumes that each node i can be seen as an oscillator with its own frequency $ \omega_{i} $ and is coupled with its neighbor through :
\[\dot{\theta_{i}}=\omega_{i} + \dfrac{K}{N}\sum_{j=1}^{N}sin(\theta_{j}-\theta_{i}+\delta)\]
where K is a constant coupling strength. This model is often rewrite in order to handle the graph topology through the adjacency matrix $ A=[a_{ij}] $:
\[\dot{\theta_{i}}=\omega_{i} + \sigma\sum_{j=1}^{n}a_{ij}sin(\theta_{j}-\theta_{i})\]
where $ \sigma $ is the coupling strength. 

The synchronization often exhibits a second phase order transition which can be seen as follow : nodes oscillate at their own frequencies for low coupling, but, as the coupling strength increases above a certain threshold, small clusters of synchronized oscillators start appearing in the network. If the coupling strength increases further, more and more oscillators lock their phases, which results in a growing order parameter. If the natural frequencies are not too spread and the coupling is large enough, the whole system may settle in a completely synchronized state. (cf. figures \ref{fig:synchro} and \ref{fig:desynchro}).

Starting from a basic equilibrium equation :
\[P_{source}=P_{dissipated}+P_{accumulated}+P_{transmitted}\] 
\cite{Filatrella2008} shows that the dynamics of a network of rotating machines (generators and loads) could be approximated by a second order Kuramoto model on the voltage phase angles of these units : 
\[\ddot{\tilde{\theta_{i}}}=W_{i}-\alpha \dot{\tilde{\theta_{i}}} + K\sum_{j\neq i}a_{ji}sin(\tilde{\theta_{j}} - \tilde{\theta_{i}})\]
where the dissipation parameter $ \alpha $, the coupling constant K, and the $ W_{i}'s $ distribution can be expressed as functions of the systems parameters.

This equation model (also known as swing equation model) has been recently employed in litterature in order to exhibit power grid synchronization properties. \cite{Motter2013} uses for instance a similar model to show that the stability of synchronous states in power grids could be enhanced by tuning some parameters of the network units. Even if topology has often been considered in previous studies as a determinant factor for the synchronization of the dynamical units, \cite{Motter2013} explains that, since the power grid cannot be reduced only at a physical network of transmission lines, the enhancement of synchronization can be explored along other leads closer to the dynamic properties of the system.

\cite{Dorfler2013} proposes a model where generators (set $ V_{1} $ of nodes) and loads ($ V_{2} $) are described by different dynamics :
\[ \left. \begin{array}{lll}
M_{i}\ddot{\theta_{i}}+D_{i}\dot{\theta_{i}}=\omega_{i}-\sum_{j=1}^{n}a_{ij}sin(\theta_{i}-\theta_{j}),\ \ i\in V_{1} \\
\ \ \ \ \ \ \ \ \ \ D_{i}\dot{\theta_{i}}=\omega_{i}-\sum_{j=1}^{n}a_{ij}sin(\theta_{i}-\theta_{j}),\ \ i\in V_{2}
\end{array}
\right. \]
Where $ M_{i} $ and $ D_{i} $ can be seen as inertia and damping coefficients for the generators while $ D_{i} $ is comparable to time constants for the loads.
\cite{Dorfler2013} intoduces then two synchronization notions :
\begin{itemize}
\item \textit{Synchronized frequencies} : $ \forall i \in V,\ \dot{\theta_{i}}=\omega_{SYNC} $, where $ \omega_{SYNC} $ is a constant common value.
\item \textit{Cohesive phases} : $ \forall (i, j) \in \xi,\ \exists \gamma \in [0; \pi /2[,\ \left| \theta_{i}-\theta_{j} \right| \leqslant \gamma $
\end{itemize}
A key point of \cite{Dorfler2013} comes from its proposed synchronization condition, which states that the previous coupled oscillator model has a unique and stable solution with respect to the previous two synchronization notions if :
\[||L^{\dagger}\omega||_{\varepsilon,\infty}\leqslant sin(\gamma)\]
Where $ L^{\dagger} $ is the Moore–Penrose pseudo-inverse of the network Laplacian matrix and $ ||x||_{\varepsilon, \infty} = max_{\{i,j\}\in \varepsilon} |x_{i}-x_{j}| $.

\begin{figure}[ht]
\centering
\includegraphics[scale=0.3]{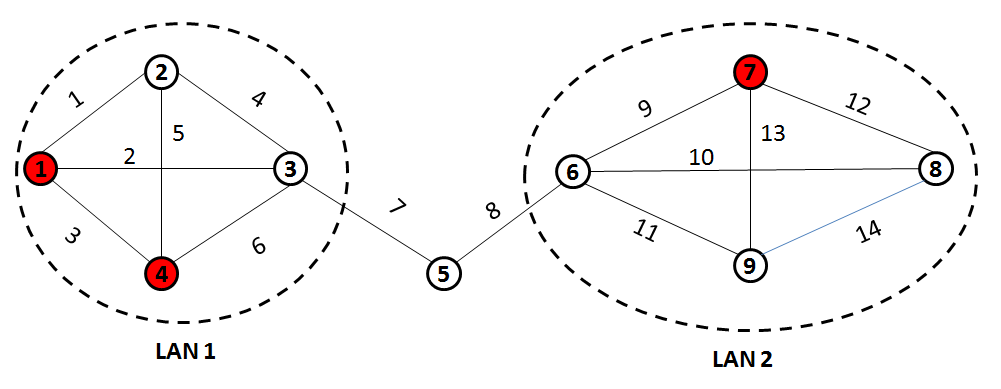}
\footnotesize{ \caption{Example of a network. Each node is an oscillator with its own natural frequency. Red nodes have positive natural frequencies while white nodes have negative ones. Edge weights $ e_{ij} $ represent the coupling strength between oscillators. (Those on the scheme are only examples) \label{fig:LAN}}}
\end{figure}

\begin{figure}[ht]
\centering
\scalebox{0.5}
{\includegraphics{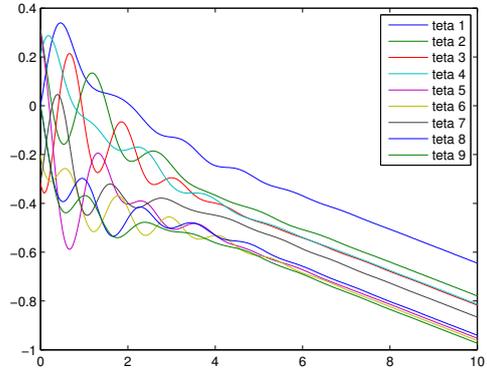}}
\footnotesize{ \caption{Synchronization of the phase angles in the oscillators network (cf. figure \ref{fig:LAN}) (coupling strengths are different than those on the scheme) \label{fig:synchro}}}
\end{figure}

\begin{figure}[h!]
\centering
\scalebox{0.5}
{\includegraphics{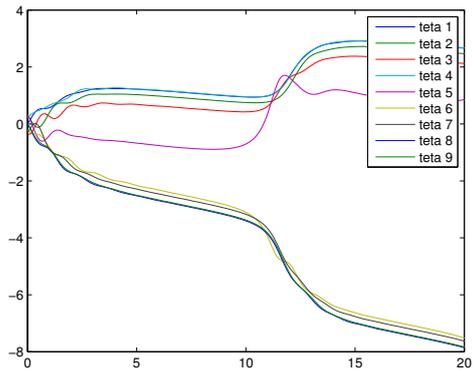}}
\footnotesize{ \caption{Here (same network (cf. figure \ref{fig:LAN}) but with more heterogeneous natural frequencies), natural frequencies are too spread and the coupling strengths are not high enough to prevent the formation of synchronized clusters without global synchrony (coupling strengths are different than those on the scheme) \label{fig:desynchro}}}
\end{figure}

These dynamical models provide interesting tools to study synchonization and stability in power grids. They also enable further researches such as determining how to maintain stability in a decentralized manner with only a restrained knowledge of the topology. For instance, in the example network in figure \ref{fig:LAN}, how could a node adjust his frequency so that the whole system remains synchronized with only a local knowledge such as neighboring or same LAN nodes states ?

\subsection{Consensus}
Since previous models fit power grids, they might also be used somehow to model units' dynamics on the physical layer of smart grid systems. However, a question of interest is whether they could be used in order to model behaviors on higher layers. The Kuramoto model can actually be seen as a particular case of consensus algorithm where oscillators try to find a consensus on a common frequency value. Considering more complex entities such as prosumers or microgrids for instance as well as other variables than frequencies, would it be possible to use consensus algorithms as distributed control strategies ?

Actually, consensus litterature is relatively wide as there are many ways of defining and handling it. Consensus is indeed often studied in the animal world in order to understand how organized group behaviors could happen whithout apparent leader \cite{Conradt2005}. Consider a well organized flock of birds for instance, flying towards a common direction and holding, as a group, a constant formation, while being populated by individuals with their own desires (direction, speed...). How can a group of units with only a neighboring knowledge agree on common things like a direction, or a rendez-vous point ?

As consensus algorithms enable dealing with distributed control, they are also well studied for various domains like engeneering or physics for instance \cite{Fax2003}. While there is a wide range of indirect utilizations (see \textit{consensus clustering} for instance \cite{Lancichinetti2012}), a classical problem is the average consensus, aiming at obtaining a common mean value of a phenomenon. Consider a network of sensors for instance, each of them picking up a local noisy value of a global phenomenon. It has been shown that through relatively simple and distributed algorithms where each units calculates the mean between its own value and its neighbors' before broadcasting the result, all units converge to the same mean value.

As the distributed notion is part of the smart grid philosophy, a few publications suggested to use algorithms inspired by consensus in order to obtain a proper decentralized control for the system. \cite{Zhang2012} introduces for instance a distributed control algorithm for solving an economic dispatch problem among generating units facing a changing demand. The consensus variable considered here is the incremental cost of each generation unit. Unlike a centralized control where the controller acquires all the information, calculates the incremental costs of each unit, and then, send the information, a decentralized approach considers that each unit has a controller that updates his own value according to its neighbors'. Nevertheless, \cite{Zhang2012} uses a leader among all the units in order to control whether to increase or decrease the group incremental cost : if the total power production is higher than the load, then the leader makes the group incremental cost decrease.

Units can be represented by their cost functions $ C_{i}(P_{Gi})=\alpha_{i}+\beta_{i}P_{Gi}+\gamma_{i}P_{Gi}^{2} $, which indicates the cost of producing $ P_{Gi} $ units of power. Assuming that generators have operating constraints $ P_{Gi,min} $ and $ P_{Gi,max} $, the objective of the economic dispatch problem is to minimize :
\[ C_{total} = \sum_{i=1}^{n}C_{i}(P_{Gi}) \]
Under constraints :
\[ \left. \begin{array}{lll} P_{D}-\sum_{i=1}^{n}P_{Gi}=0 \\
                             P_{Gi,min}\leqslant P_{Gi}\leqslant P_{Gi,max}                      
                             \end{array}
                             \right. \]
                             Where $ P_{D} $ represents the total power demand.
It is then possible to define the natural incremental costs $ \lambda_{i} $ of each unit based on its cost fuction :
\[\lambda_{i} = \dfrac{\partial C_{i}(P_{Gi})}{\partial P_{Gi}}\ \ \ \ \forall i \in [1,N] \]
Assuming a sudden increase in power demand for instance, generators of the network have to adapt their outputs in the most effective way (here represented by the minimization of the total cost function), so that the total output sum matches the demand. Each unit sends then its incremental cost to its neighbors and updates its own according to the following first-order consensus algorithm :
\[ \left\{ \begin{array}{lll} \lambda_{i}[k+1] = \sum_{j=1}^{n}d_{ij}\lambda_{j}[k] + \epsilon \Delta P\ \ for\ the\ leader \\
                             \lambda_{i}[k+1] = \sum_{j=1}^{n}d_{ij}\lambda_{j}[k]\ \ \ \ \ \ for\ the\ followers 
                             \end{array}
                             \right. \]
Where $ \Delta P = P_{D}-\sum_{i=1}^{n}P_{Gi} $ represents the mismatch between the total demand and the overall power generated, and $ \epsilon $ is a convergence parameter. The $ d_{ij}'s $ are the elements of the row-stochastic matrix D and can be defined based on the network Laplacian elements $ l_{ij} $ :
\[d_{ij}= \dfrac{|l_{ij}|}{\sum_{j=1}^{n}|l_{ij}|}\]

\begin{figure}[ht]
\centering
\includegraphics[scale=0.4]{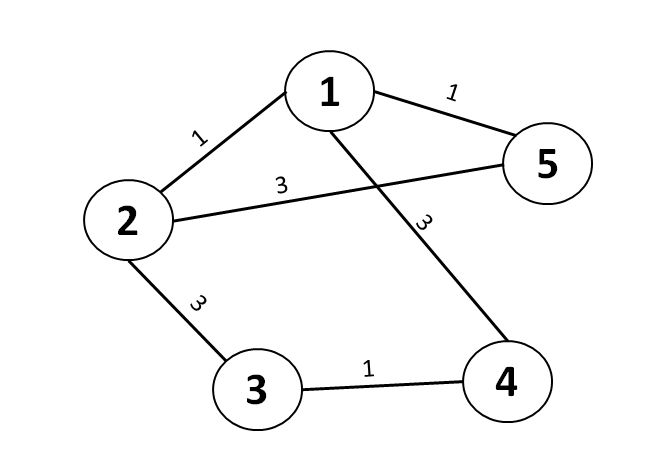}
\footnotesize{ \caption{Example of a simple communication network between generators controllers. Edge weights $ e_{ij} $ represents time-delays (numbers on the figure are only examples), ie  $ \tau_{ij}=e_{ij}\Delta t $. ($ \Delta t $ being the time step.) \label{fig:network}}}
\end{figure}

\begin{figure}[h!]
\centering
\scalebox{0.5}
{\includegraphics{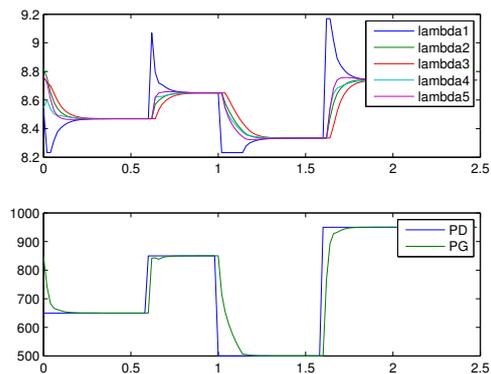}}
\footnotesize{ \caption{Leader consensus in a 5 units (figure \ref{fig:network}) network without time delays and under power demand variations. The first chart represents the evolution of the $ \lambda_{i}'s $ while the second shows the evolution of power demanded and generated} \label{fig:sans}}
\end{figure}

\begin{figure}[h!]
\centering
\scalebox{0.5}
{\includegraphics{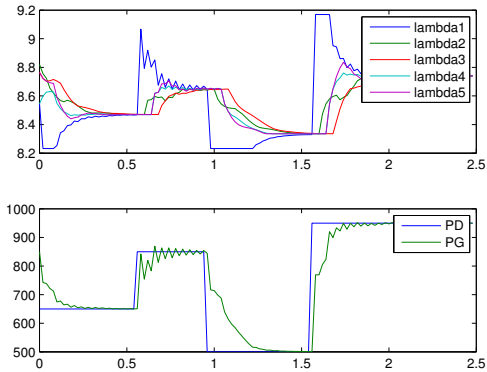}}
\footnotesize{ \caption{Leader consensus in the same 5 units network (cf. figure \ref{fig:network}) with time delays and under power demand variations. The first chart represents the evolution of the $ \lambda_{i}'s $ while the second shows the evolution of power demanded and generated.  \label{fig:avec}}}
\end{figure}

\begin{figure}[h!]
\centering
\scalebox{0.5}
{\includegraphics{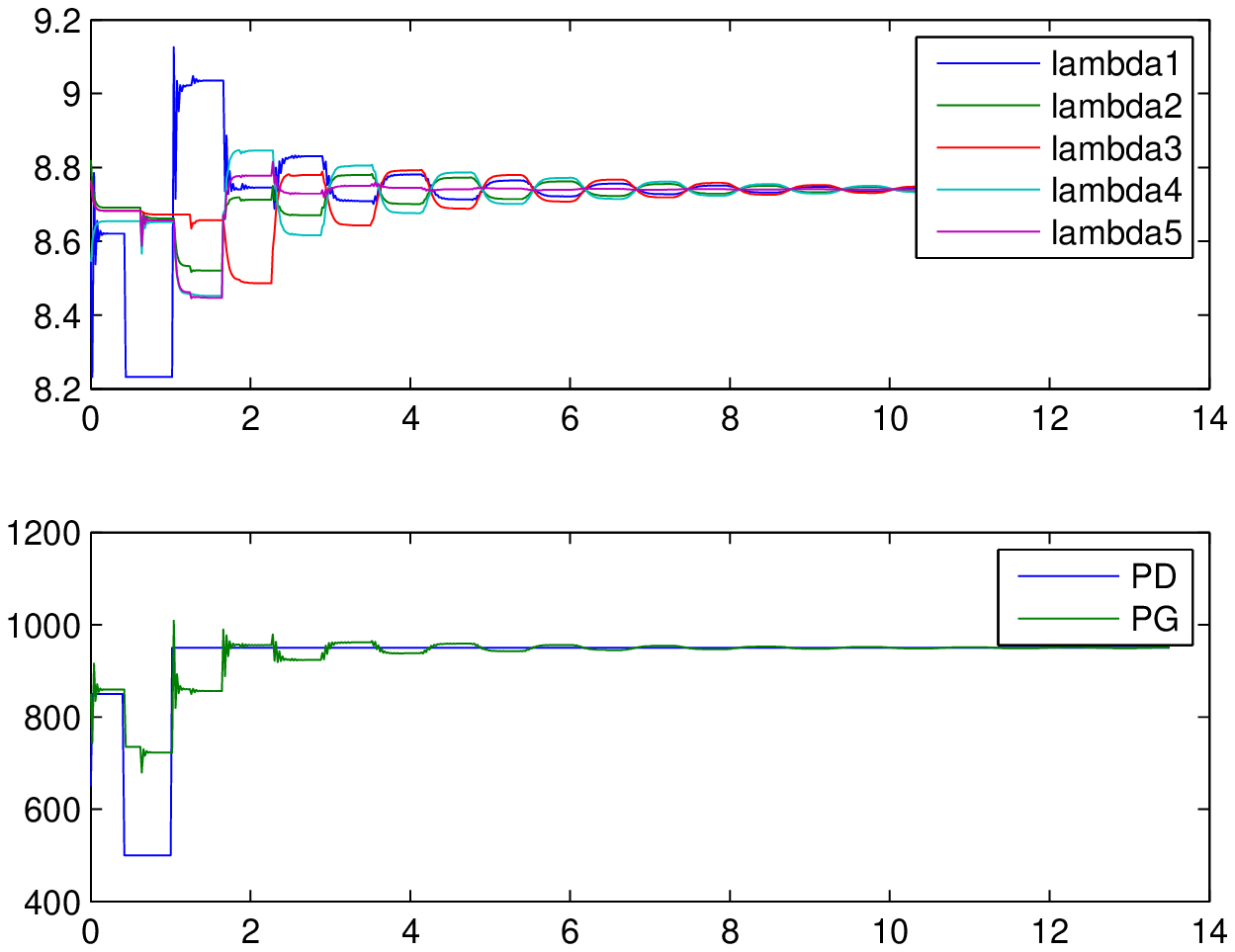}}
\footnotesize{ \caption{Leader consensus in the same 5 units network  (cf. figure \ref{fig:network}) with important time delays and under power demand variations. The first chart represents the evolution of the $ \lambda_{i}'s $ while the second shows the evolution of power demanded and generated. The system oscillates but converges asymptotically toward a consensus value \label{fig:gros}}}
\end{figure}

(cf. figures \ref{fig:sans}, \ref{fig:avec}, and \ref{fig:gros} for simulation results in a 5 units network (figure \ref{fig:network})).

Hence, network topology is a key parameter in the convergence speed of this distributed algorithm. Actually, convergence speed of consensus algorithms is a question of interest and is frequently tackled in litterature. It has been shown that performance of consensus algorithms could be linked to the algebraic connectivity of the graph $ \lambda_{2} $ (also called the Fiedler eigenvalue), which is basically the smallest non-zero eigenvalue of the underlying graph's Laplacian. More precisely, the group disagreement vector $ \delta $ globally asymptotically vanishes with a speed equal to $ \lambda_{2} $ :
\[||\delta(t)|| = ||\delta(0)||exp(-\lambda_{2}t)\]
Indicating that the larger $ \lambda_{2} $, the faster the convergence will be. Or, in less precise words, the more connected the network is, the faster a consensus will be reached. 

\begin{figure}[h!]
\centering
\scalebox{0.5}
{\includegraphics{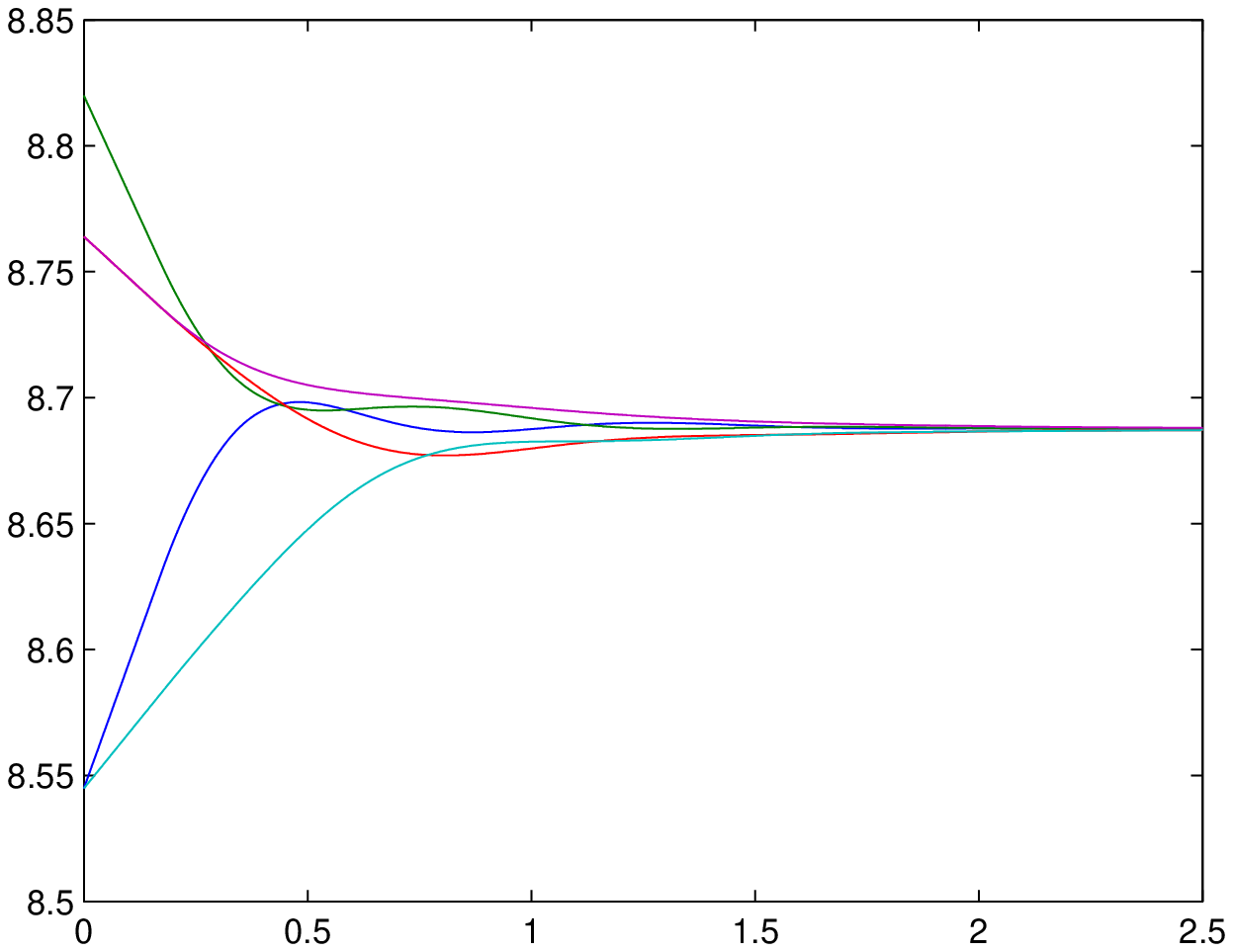}}
\footnotesize{ \caption{Leader-less consensus in figure \ref{fig:network} network under equal communication time-delays $ \tau=\tau^{\star}/2 $. }  \label{fig:oscconv}}
\end{figure}

\begin{figure}[h!]
\centering
\scalebox{0.5}
{\includegraphics{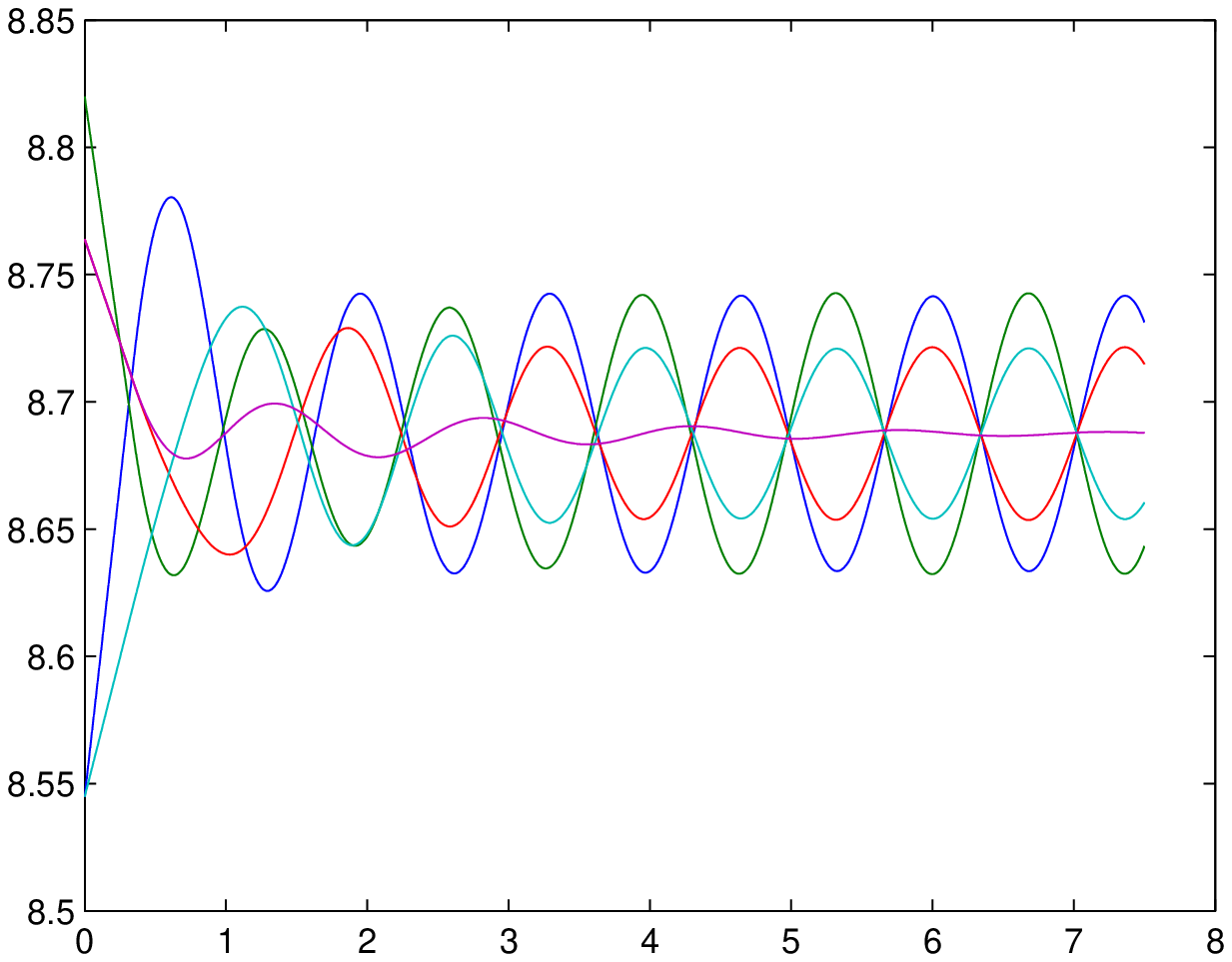}}
\footnotesize{ \caption{Leader-less consensus in figure \ref{fig:network} network under equal communication time-delays $ \tau=\tau^{\star} $.} \label{fig:osc}}
\end{figure}

\begin{figure}[h!]
\centering
\scalebox{0.5}
{\includegraphics{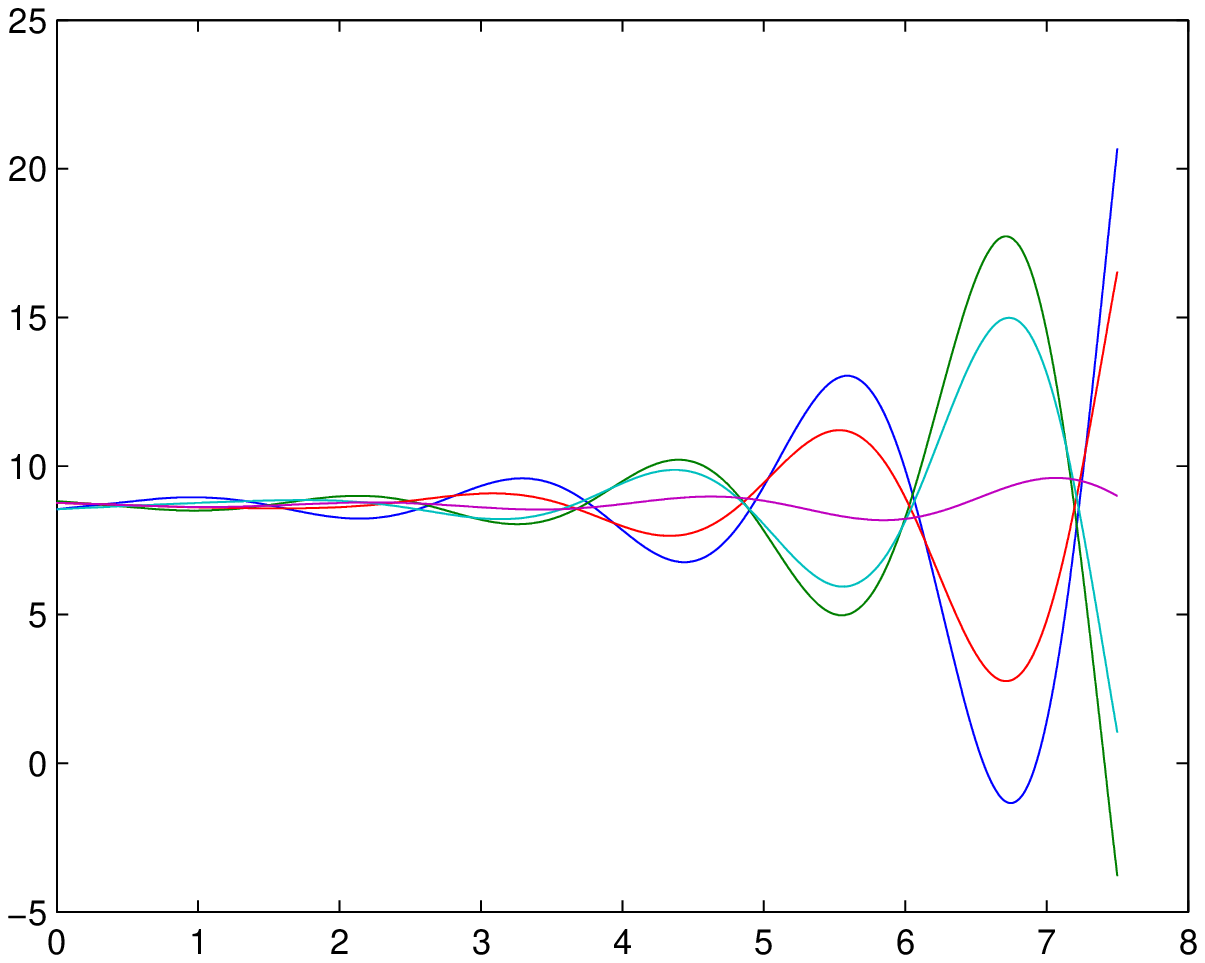}}
\footnotesize{ \caption{Leader-less consensus in figure \ref{fig:network} network under equal communication time-delays $ \tau=2 \tau^{\star} $.} \label{fig:oscdiv}}
\end{figure}

In order to be more realistic, variants of these consensus algorithms have been introduced such as consensus in directed graphs, switching topology networks, communication time-delays networks... \cite{Olfati-Saber2004} \cite{Olfati-Saber2007}
In the case where time-delays are introduced, the convergence performances are not as obvious as the simple previous case. Actually, it has been shown (\cite{Olfati-Saber2004}) that there exists a trade-off between performance and robustness, meaning that networks with hubs have difficulties to tolerate high communication time-delays.
More precisely, consider the following consensus algorithm including time-delays :
\[ \dot{x_{i}}(t) = \sum_{j=1}^{N}a_{ij}[x_{j}(t-\tau_{ij})-x_{i}(t-\tau_{ij})]\]
Under the assumptions that the network is fixed, undirected, connected, and that all time-delays are equals ($ \forall (i,j),\ \tau_{ij}=\tau $), the previous algorithm globally asymptotically solves the average-consensus problem if and only if one of the following conditions is satisfied :

\begin{itemize}
\item $ \tau \in (0,\tau^{\star}) $ with $ \tau^{\star}=\pi / 2\lambda_{n} $ ($ \lambda_{n} $ being the maximum eigenvalue of the network Laplacian matrix). (cf. figures \ref{fig:oscconv} \ref{fig:osc} \ref{fig:oscdiv}).
\item The Nyquist plot of $ \Gamma(s)=e^{-\tau s}/s $ has a zero encirclement around $ -1/\lambda_{k},\ \forall k > 1 $
\end{itemize}

Furthermore, for a time-delay equals to $ \tau^{\star} $ the system oscillates asymptotically with a frequency equals to $ \lambda_{n} $ (cf. figure \ref{fig:osc}).

A direct consequence is that, by scaling down the edge weights, an arbitrary delay could be tolerated : $ \forall \tau > 0,\ \exists ! k > 0,\ \tau < \pi / (2k\lambda_{n}) $. However, this has the consequence of reducing $ \lambda_{2} $ by a $ 1/k $ factor, meaning that the performance on the convergence is altered. There is thus a trade-off between performance and robustness against time-delays that have to stay in a $ [0, \tau^{\star}] $ window for the system to reach a consensus, which is different from the leader protocol introduced by \cite{Zhang2012}. Actually, in the leader consensus algorithm (under the assumptions of \cite{Zhang2012}), agents converge asymptotically to the consensus value for all delay values. That is, in this (noise-less) leader consensus algorithm, a consensus is always asymptotically reached as long as the network is connected (cf. figures \ref{fig:sans}, \ref{fig:avec}, and \ref{fig:gros}). 

Nevertheless, as explained in \cite{Zhang2012}, the leader consensus protocol implies the election of a leader based on criteria such as the highest betweeness centrality for instance, which induce interesting questions on a possible decentralized protocol for electing the most efficient leader of a group.

\section{Smart grid decentralized control under telecommunication reality}

Consensus decentralized algorithms are interresting tools toward control models. However, a proper control algorithm for the smart grid system should be designed according to the reality of the infrastructure it will be implemented on. 

Let's consider for example that a consensus algorithm, similar to the one discussed avove, is responsible for the attribution of generators outputs. Incremental costs information are then exchanged through a communication network (latencies, packet losses, queing, errors). All of these notions are part of best-effort data networks and will, with no doubt, impact in return the power grid. Imagine for instance that the packets transporting the incremental costs information of the example considered above are lost or delayed. This means that generators will use outdated information to adapt their outputs, which lead to a production unadapted to consumption, which can in turn lead to stability problems, and finally results in black-outs.

The whole smart grid project is seen as the opportunity to modernize and attribute new applications to the power grid. Nevertheless, there is a gap between complex application ideas and the reality, meaning that physical constraints will in the end seperate what is feasible from utopian conceptual ideas. Building and studying models is the first step towards establishing engineering rules for the system and obtaining a proper normalization for the smart grid. 

One of the main actors in this field is the american National Institute of Sciences and Technology (NIST), which formed working groups to study smart grid applications' requirements in terms of telecommunication specifications (cf. NIST smart grid priority actions PAP). This may well only be a portion of the whole system, the scope remains still very wide as there are numerous possible action points.

Communication networks are indeed multi-layered themselves (cf. OSI layers) and engineers have established various and very different protocols for each layer. Obviously, some of them may not be adapted for the smart grid communication system, while others might need modifications. Consider the widely used transport/network layers combination TCP/IP for instance. According to \cite{Bakken2011}, using TCP for sensors' synchrophazors transport might not be adapted as TCP re-transmit packets until receiving acquitments. To be usefull for control, synchrophazors from various points of the grid have to be time-coherent (have the same timestamp), meaning that if a synchrophazor information is lost, it is worthless to re-transmit it. Even worse, this might congestion the data network even more.

Of course, this small example doesn't mean that TCP is not adapted for the smart grid telecommunication nework as it might be usefull for the transport of other applications than synchrophazors management. It however gives a glimpse of the complexity of finding the network requirements per application in terms of protocols, topology, physical infrastructure (wireless, optical fiber...), and physical constraints (latency, losses...), a still very recent approach of the smart grid.

\section{Perspectives and conclusion}
Through this survey, we decorticate the smart grid system in order to focus on the communication part while keeping in mind that it has to be coherent with the rest of the system. In other words, depending on the section studied, different requirements towards other portions are likely to appear. A complex task revolves around specifying these requirements as well as their tolerances, especially when really complex concepts such as virtual power plants or super islanding are considered.

Such a system cannot be built overnight, and a few preliminary steps are necessary. NIST is currently searching and proposing communication network requirements for smart grid applications such as large scale smart metering management. We believe that building informatic large scale models based on complex system theory tools and incorporating such data might enable us to re-adjust some of the requirements and study how the system reacts against stress and perturbations. 

High level concepts such as virtual power plants and islanding obviously call for strong constraints on the communication network. If such behaviors are enabled, it is likely that some established parameters will have to change. It is a question of interest to study, first through modeling, wether these concepts could be someday implemented in a large scale smart grid system and what would their consequences be in terms of stability and resilience.

\listoffigures

\footnotesize{
\bibliographystyle{plain}
\bibliography{survey1}
}
\end{document}